\documentclass[aps,preprint]{revtex4}
\usepackage{amsmath}
\usepackage{amssymb}
\usepackage{graphicx}

\begin{document}

\title{Incorporation of the intensive and extensive entropy
contributions in the disk intersection theory of a hard
disk system}
\author{V.M. Pergamenshchik$^{1,2}$ }
\email{victorpergam@cft.edu.pl}
\affiliation{$^{1}$Institute of Physics, National Academy of Sciences of Ukraine,
prospekt Nauky, 46, Kyiv 03039, Ukraine \\
$^{2}$Center for Theoretical Physics, Polish Academy of Sciences, Al. Lotnik%
\'{o}w 32/46, Warsaw, Poland \\
}
\date{\today }

\begin{abstract}
The one-body free volume, which determines the entropy of a hard disk
system, has extensive (cavity) and intensive (cell) contributions. So far
these contributions have not been unified and considered separately. The
presented theory incorporates both contributions, and their sum is shown to
determine the free volume and partition function. The approach is based on
multiple intersections of the circles concentric with the disks but of twice
larger radius. The result is exact formulae for the extensive and intensive
entropy contributions in terms of the intersections of just two, three,
four, and five circles. The method has an important advantage for
applications in numerical simulations: the formulae enable one to convert
the disk coordinates into the entropy contribution directly without any
additional geometric construction. The theory can be straightforwardly
applied to a system of hard spheres.
\end{abstract}

\maketitle

\section{Introduction.}

Spheres and their arrangement in space has been playing a very notable role
in practical life and, in particular, in mathematics and physics. Recently
Maryna Vjazovska received Fields medal for solving the problem of dense hard
sphere packing in 8 \cite{Maryna8} and 24 \cite{Maryna24} dimensions. In
physics, however, the analytical achievements are more modest. In
statistical physics we are interested in a random packing problem of hard
spheres in a macroscopic volume. Over more than a century, the idea to model
molecules as hard spheres has been widely used in the theory of liquids \cite%
{JCP1,JCP2,JCP3}. The model of spheres, interacting only by their hard
cores, plays the role similar to that of Ising's model in the theory of
magnetism, but despite apparent simplicity, the behavior of hard sphere
systems is so complex mathematically that no exact analytical result has
been obtained in the physical dimensions 3 and 2. Under these circumstances,
the numerical Monte Carlo and molecular dynamics approaches have become the
main tools in the study of 3D hard sphere and 2D hard disk (HD) systems (see
review \cite{JCP2021} and numerous references therein). However even using
the modern powerful numerical methods one encounters the fundamental
theoretical problem of computing the main thermodynamic potential of hard
sphere and HD systems, the entropy. Although the problem of hard spheres is
very similar to that of HDs, for simplicity, in this paper the presentation
will be mainly related to HDs.

The potential energy of a HD system is zero and the entropy gives the total
thermodynamic information and, in particular, equation of state and possible
nuances of the phase behavior as a function of density which is the single
parameter of the system state. The numerical simulations consist of
producing different independent configurations of disks' coordinates which
is the task input, and then certain related theory must provide calculation
of the entropy, equation of state, and other quantities of interest which
are the task outcome. In principle, the ultimate theory must give the
outcome directly from the disks' coordinates (with the consequent averaging
over different configurations), but actually the available theoretical
methods relate disks' coordinates with the expected outcome only through
intermediate and quite sophisticated geometrical constructions. The main
elements of such constructions are the so-called free volume, cavity,
"private" one-disk cell, and the surface thereof. For a brief review and
presenting the main idea of this paper, I first introduce these quantities.

A HD of radius $\sigma /2,$ the core, is supplemented with a concentric
circle of radius $\sigma $ which is called here $\sigma $ circle. The cores
cannot overlap, but their $\sigma $ circles can and are transparent for
cores. A configuration of $N$ HDs in a 2D volume $V$ consists of $N$
nonoverlapping cores and $N$ connected $\sigma $ circles which can overlap,
Fig.1. The \emph{free volume }$V_{N}$ of a disk\emph{\ }in a given
configuration of an \emph{equilibrium }$N$\emph{\ HD system} is the volume
accessible for its center in this configuration, which is the total $V$
minus union of the rest $N-1$ $\sigma $ circles; $V_{N}$ can comprise more
than one disconnected pieces (in Fig.1, \ a single piece is shown). The 
\emph{cavity }$C_{N}$\emph{\ in a HD system of }$N$\emph{\ HDs} is the area
where an additional HD can be inserted which is total $V$ minus \emph{union
of all }$N$\emph{\ }$\sigma $\emph{\ circles}. The private cell $c_{N}$ of a
disk \emph{in a HD system of }$N$\emph{\ HDs} is the free volume of this
disk $V_{N}$ without the cavity $C_{N}$ (stroked area), Fig.1; if the cavity
is zero and no additional HD can be inserted, then private cell is the total
free volume of the disk. The free volume, cavity, and private cell in the
case of a single free area are illustrated in Fig.1. It is seen that the
division on cavity and private cell of \ the dashed disk depends on the
position of this disk, but the total free volume $V_{N}=C_{N}+c_{N}$ does
not depend on its coordinate. The average free volume $\left\langle
V_{N}\right\rangle _{N},$ average cavity $\left\langle C_{N}\right\rangle
_{N},$ and average private cell per disk $\left\langle c_{N}\right\rangle
_{N}$ are those for a single configuration of $N$ disks averaged over the
configurations of the \emph{equilibrium system of same }$N$\emph{\ disks. }%
The perimeters (surfaces in 3D) of all the three volumes introduced are
complex lines (surfaces) whose shapes and lengths are not in a one-to-one
relation with the volume size. In the above definition, we emphasized that
the quantities related to an $N$ disk system are defined for the equilibrium
system of same $N$ disks.

In 1977 Speedy introduced the spare volume and defined it like that:
\textquotedblleft The spare volume $V_{s}$ of an assembly of $N$ spheres of
diameter $\sigma $ in a volume $\ V$ is defined as the average over
configurations of the volume which is not within $\sigma $ of a sphere
center,..., the probability that another sphere can be placed at a random
point in the assembly\textquotedblright\ \cite{Speedy77}. Clearly, this is
equivalent to a cavity available for an additional, $N+1$ st disk in a
system of $N$ disks, and which is averaged over configurations of $N$ HDs.
As a result, Speedy related the partition function (PF) of a HD system with
the product of cavities $\left\langle C_{N^{\prime }}\right\rangle
_{N^{\prime }-1}$ in the systems of number of disks reduced by one, i.e., of 
$N-1,$ $N-2,...,1,$ $0$ disks, which are averaged over the equilibrium
systems of respectively $N-1,$ $N-2,...1,$ $0$ disks in the same volume $V$ 
\cite{Speedy77}. The way this PF was obtained was going back to the earlier
results by Adams \cite{Adams} and Anrews \cite{Andrews} which had in turn
been inspired by Widom's approach \cite{Widom}. But Speedy was the first to
address the calculation of the spare volume in a HD system in terms of
disks' intersections \cite{Speedy80} which has greatly influenced the
further development of this area \cite{Melnyk}. Later Speedy explicitly
shifted from the nomenclature of spare volumes to cavities \cite{Speedy80}.

Thus, Speedy approach does not incorporate the free volumes but only their
parts, cavities. Actually, however, the formula for the PF has not been
further employed. Instead, in 1980 \cite{Speedy80,Speedy81} Speedy proposed
the equation of state which relates the pressure with the ratio (average
cavity volume)/(average cavity surface area). Since then different
geometrical methods of finding the cavities and their surface have become
the main emphasis in the ongoing studies of hard particle systems \cite%
{1995,1997,1998,1999,2006,2014,2015,2016}. However, the complex shape and
connectedness of cavity space make precise measurement of the quantities,
characterizing them, very difficult which resulted in new and new
geometrical constructions that are highly nontrivial to implement \cite%
{1995,1997,1998,2014,2015}. The main problem of this approach has been that,
even for densities far from the crystallization density, cavities become so
rare that finding them was sometimes called a task futile \cite{1998}. The
root of this problem is that a cavity in an $N$ HD system has been mainly
computed as that of $N$ th disk in an equilibrium system of $N-1$ HDs. But 
whereas in an equilibrium system of $N$ HDs the place for $N$ th disk is
guarantied, in a dense equilibrium system of $N-1$ HDs, a place for an
additional $N$ th disk is a very rare event. The paradox is that relating
cavity with a system of a smaller number of disks when considering an $N$
disk system, one finds no place for $N$ th disk. This problem was pointed
out by Schindler and Maggs who had to invent a modified numerical algorithm
for finding cavities and distinguishing them from free volumes \cite{2015}.

The cavity is an extensive quantity that scales with the number of disks and
volume. At the same time, even when an additional, $N+1$ st disk cannot be
inserted, the original $N$ disks can vibrate in their cages created by their
neighbors. This implies that the total entropy is nonzero and the volumes of
such cages are its source even in the absence of cavities which is the case
of densities approaching that of crystallization. These cages are what is
called private disk cells in the cell models \cite{1968,1972,1979,PRE2006}.
Even before Speedy's publication \cite{Speedy77}, Hoover and coworkers \cite%
{1968,1972,1979} correctly argued that along with the extensive cavity
volume, there must be an intensive one which scales as $V/N$ and consists of
individual single-disk cells. The free volume is the sum of these two terms,
and when the cavity is getting smaller and smaller, the total free volume
reduces to the volume of individual cells. Based on this important idea, as
early as in 1972, Hoover, Ashurst, and Groover showed that the pressure can
be expressed via the ratio of the average of the\emph{\ free volume} to the
average of its surface which incorporates the cell contribution \cite{1972}.
The cell model can quantitatively describe the HD equation of state near
freezing density in numerical simulations \cite{1968,1972,1979} and even
allows one to obtain qualitatively accurate results analytically \cite%
{PRE2006}. The cell models also have the problem of describing the cell
distribution with a strong geometric component, but the main problem is to
connect the one-body cell approach with a many-body one, i.e., to
incorporate the intensive and extensive free volumes and entropy
contributions in a unified theory.

In this paper I present such a theory in which the extensive and intensive
terms have the same status and are computed in the framework of the same
approach. I develop the method of multiple intersections of $\sigma $
circles and, in terms of their intersection volumes, express the free
volume, its extensive and intensive parts, and the PF of a system of $N$
HDs. Due to the fact that only up to five $\sigma $ circles can intersect
without overlapping of their cores, the theory needs only four quantities,
i.e., the intersection volumes of two, three, four, and five $\sigma $
circles. These four quantities are fully specified by the disks coordinates
and can be calculated analytically using formulae obtained in \cite%
{Kratky1,Kratky2, Sheraga}. The theory does not resort to a system of
reduced number of HDs and gives the values of a cavity, private cell, and
total free volume in a system of $N$ HDs only in terms of this very $N$ HD
system. No geometrical or any other intermediate constructions appear
between the input, disks' coordinates, and the output, quantities of
interest, and the only source of inaccuracy is that of the numerical
simulations.

The paper is organized as follows. In Sec.2, the method of multiple disk
intersections is introduced and the general formula for the free volume of a
single disk is derived. Sec. 3 is devoted to the connection between the
single disk volume and many-body description. First Speedy-Widom's approach
is used to derive the PF. It is shown that this PF is exactly Speedy's PF 
\cite{Speedy80} in the form of product of cavities in the equilibrium
systems of the reduced number of disks. Next it is shown that, in the
thermodynamic limit, the correct PF is the product of the free volumes
averaged in the proper systems. In Sec.4, the formulae relating the
extensive and intensive free volume contributions with intersections of $%
\sigma $ circles are obtained and their application to the average values is
explained. In Sec.5, the analytical computation of all the intersections of $%
\sigma $ circles and the intensive and extensive terms for the densely
packed triangular HD lattice is presented in detail. It shows that both
terms in this state vanish identically. Final Sec.6 is a brief conclusion. 
\begin{figure}[tbp]
\includegraphics[height=10cm]{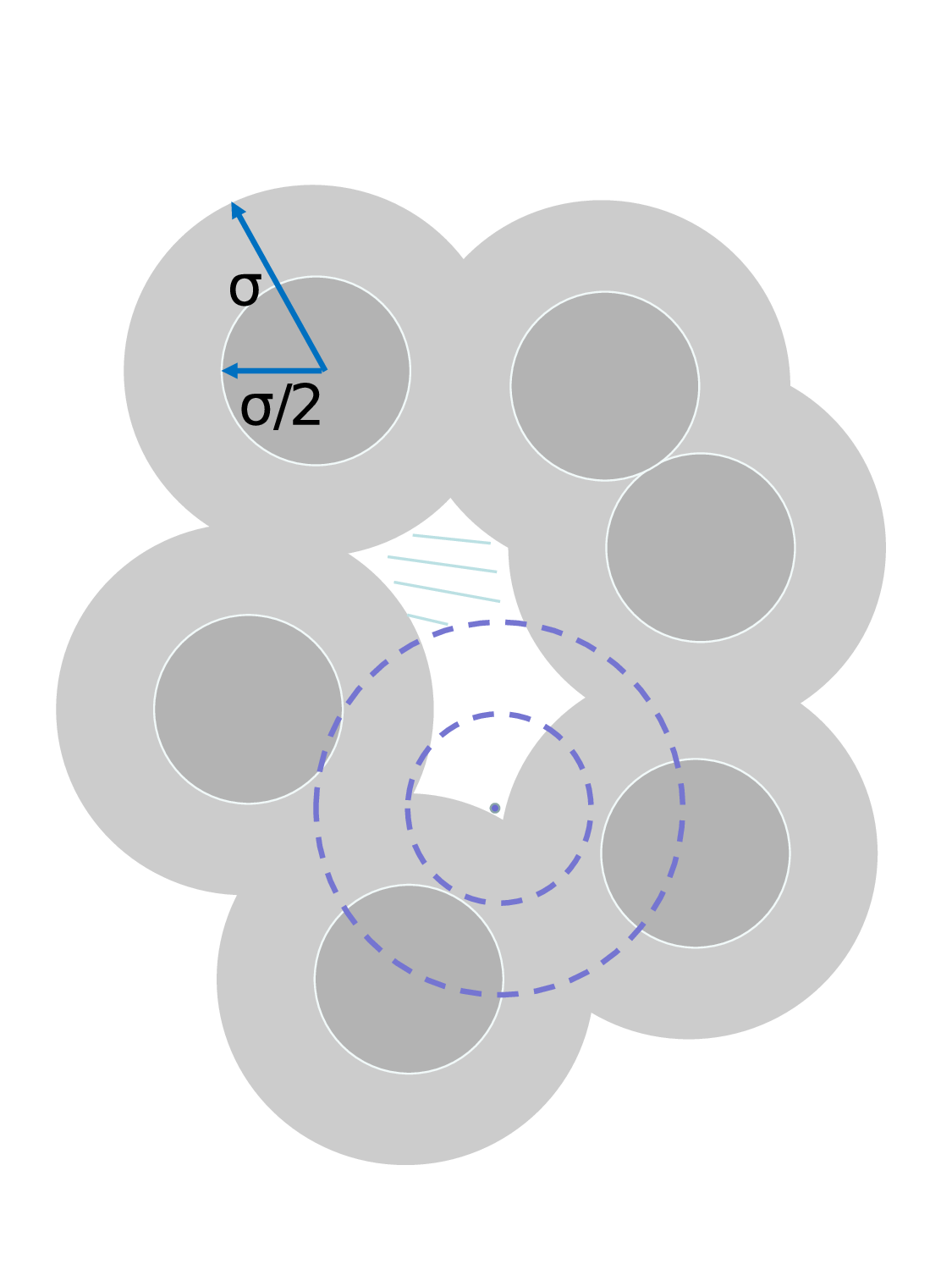}
\caption{{}Fragment of a system of $N$ HDs. The $N-1$ HDs of radius $\protect%
\sigma /2$ are represented by dark circles and the connected concentric $%
\protect\sigma $ circles by light circles. The $N$ th disk and circle are
shown by dashes. The white area in between is the free volume of $N$ th disk
as its center can be anywhere in this area. Only the stroked fraction of the
white area is the cavity in the $N$ HD system. The clear fraction of the
white area is not cavity, but is available for the center of dashed disk
center. This clear white area is equal to the area of dashed $\protect\sigma 
$ circle, $\protect\pi \protect\sigma ^{2},$ minus its fraction overlapped
by other $\protect\sigma $ circles.}
\label{Fig1}
\end{figure}

\section{Hard disk interaction and multiple disk intersections.}

The configuration PF of a 2D system of $N$ particles in the 2D volume $V$
with paiwise interaction $U_{ij}$ is the following integral:%
\begin{eqnarray}
Z_{N} &=&\int_{V}dx^{N}\exp \left( -\frac{1}{2}\sum_{i,j=1}^{N}U_{ij}\right) 
\label{ZN} \\
&=&\int_{V}dx^{N-1}\exp \left( -\frac{1}{2}\sum_{i,j=1}^{N-1}U_{ij}\right)
\int_{V}dx_{N}\exp \left( -\frac{1}{2}\sum_{i=1}^{N-1}U_{Ni}\right) ,  \notag
\end{eqnarray}%
where $x_{i}$ are the two component vectors of disks' coordinates, $%
dx^{N}=dx_{1}...dx_{N},$ and we separated the $x_{N}$ integral. For HDs of
the radius $\sigma /2$, the potential $U_{ij}=\infty $ for $x_{j}$ within
the circle of radius $\sigma $ centered at $x_{i},$ and $U_{ij}=0$ for $x_{j}
$ outside this circle. We introduce a circle $B_{i},$ $i=1,...,N-1:$%
\begin{equation}
B_{i}=\{x_{N}:\left\vert x_{i}-x_{N}\right\vert \leq \sigma \}.  \label{Bi}
\end{equation}%
By definition, the indicator $\tau _{i}(B_{i})$ of the set of points $%
x_{N}\in B_{i}$ is 
\begin{equation}
\tau _{i}=\tau (B_{i})=\left\{ 
\begin{array}{c}
1,x_{N}\in B_{i}, \\ 
0,x_{N}\notin B_{i}.%
\end{array}%
\right.   \label{tau}
\end{equation}%
The product of $n$ indicators of $n$ different sets is the indicator of the
intersection set shared by all of them. Such product of two indicators,
which is nonzero only if the two related circles intersect, can be defined as%
\begin{equation}
\tau _{tj}=\tau _{i}\tau _{j}=\left\{ 
\begin{array}{c}
1,x_{N}\in B_{i}\cap B_{j}, \\ 
0,x_{N}\notin B_{i}\cap B_{j}, \\ 
0,B_{i}\cap B_{j}=\varnothing .%
\end{array}%
\right.   \label{titj}
\end{equation}%
Then, by definition, 
\begin{eqnarray}
\tau _{i_{1}}..._{i_{n}} &=&\tau \left(
\bigcap\limits_{k=1}^{n}B_{i_{k}}\right)   \label{titj1} \\
&=&\left\{ 
\begin{array}{c}
1,x_{N}\in \bigcap\limits_{k=1}^{n}B_{i_{k}}, \\ 
0,x_{N}\notin \bigcap\limits_{k=1}^{n}B_{i_{k}}, \\ 
0,\bigcap\limits_{k=1}^{n}B_{i_{k}}=\varnothing ,%
\end{array}%
\right.   \notag
\end{eqnarray}%
where $\bigcap_{k=1}^{n}B_{i_{k}}$ is the set of points $x_{N}$ shared by
all circles $B_{i},$ i.e., their intersection.

Now the utmost right exponential in $Z_{N}$ (\ref{ZN}) for the HD
interaction can be presented in terms of $\tau $'s. It is easy to see that
the HD interaction is equivalent to the following formula:%
\begin{equation}
e^{-U_{Ni}/2}=1-\tau _{i}.  \label{1-tau}
\end{equation}%
This formula shows that the center of any other HD $j\neq i$ cannot enter
the circle $B_{i}$ centered at $x_{i}$ which has the radius $\sigma $ twice
the HD radius $\sigma /2$. The product of two exponentials is%
\begin{equation}
e^{-(U_{Ni}+U_{Nj})/2}=(1-\tau _{i})(1-\tau _{j}).  \label{2-tau}
\end{equation}%
Similarly,%
\begin{equation}
\exp \left( -\frac{1}{2}\sum_{i=1}^{N-1}U_{Ni}\right)
=\prod\limits_{i=1}^{N-1}(1-\tau _{i})  \label{exp}
\end{equation}%
\begin{equation*}
=1-\sum_{i=1}^{N-1}\tau _{i}+\sum_{i>j}^{N-1}\tau _{i}\tau
_{j}-\sum_{i>j>k}^{N-1}\tau _{i}\tau _{j}\tau
_{k}+...+(-1)^{N-1}\sum_{i_{1}>i_{2}>...>i_{N-1}}^{N-1}\tau _{i}...\tau
_{i_{N-1}}.
\end{equation*}%
It is well-known that more than five circles of a radius $\sigma $ cannot
intersect without intersection of their cores of radius $\sigma /2$ (six HDs
intersect at a single point). As a result, all the products of six and more $%
\tau $'s do not contribute to the above sum. Thus, the last term in the sum (%
\ref{exp}) is $\tau _{ijklm}$ which corresponds to the intersection of five
circles $B$, and this formula greatly simplifies:%
\begin{equation}
\exp \left( -\frac{1}{2}\sum_{i=1}^{N-1}U_{Ni}\right) =  \label{exp tau}
\end{equation}

\begin{equation*}
=1-\left( \sum_{i=1}^{N-1}\tau _{i}-\sum_{i>j}^{N-1}\tau
_{ij}+\sum_{i>j>k}^{N-1}\tau _{ijk}-\sum_{i>j>k>l}^{N-1}\tau
_{ijkl}+\sum_{i>j>k>l>m}^{N-1}\tau _{ijklm}\right) .
\end{equation*}%
Denote by $\mu $ the volume (i.e., the measure in the 2D space, surface
area) of a set $:$ $\mu _{i}=\mu (B_{i})=\pi \sigma ^{2},$ $\mu _{ik}=\mu
(B_{i}\cap B_{j}),$ $\mu _{i_{1}...i_{n}}=\mu \left(
\bigcap_{k=1}^{n}B_{i_{k}}\right) $. Then the last integral in $Z_{N}$ (\ref%
{ZN}) reduces to the following form:%
\begin{equation}
V_{N}(x_{1},...,x_{N-1})=\int_{V}dx_{N}\exp \left( -\frac{1}{2}%
\sum_{i=1}^{N-1}U_{Ni}\right)  \label{V}
\end{equation}%
\begin{equation*}
=\theta _{N}(x_{1},...,x_{N-1})(V-V_{excl}),
\end{equation*}%
where 
\begin{equation}
V_{excl}(x_{1},...,x_{N-1})=\sum_{i=1}^{N-1}\mu _{i}-\sum_{i>j}^{N-1}\mu
_{ij}+\sum_{i>j>k}^{N-1}\mu _{ijk}-\sum_{i>j>k>l}^{N-1}\mu
_{ijkl}+\sum_{i>j>k>l>m}^{N-1}\mu _{ijklm}.  \label{Vexcl}
\end{equation}%
To exclude configurations $x_{1},...,x_{N-1}$ in which there are disks whose
hard cores overlap, we introduced the hard core indicator $\theta
_{N}(x_{1},...,x_{N-1})$ in a configuration $x_{1},...,x_{N-1}$: $\theta
_{N}=1$ if $\left\vert x_{i}-x_{j}\right\vert \geq \sigma $ for all $1\leq
i<j\leq N-1$ and $\theta _{N}=0$ if $\left\vert x_{i}-x_{j}\right\vert
<\sigma $ at least for one pair $i<j$. The quantity $%
V_{N}(x_{1},...,x_{N-1}) $ is the integral over all possible locations $%
x_{N} $ of $N$ th disk in the system of $N$ HDs for given fixed positions of
the rest $N-1$ HDs. In other words, this is the integral over the volume
accessible to the $N$ th disk in the $N$ HD system, the $N$ th disk's free
volume. It is expected to be the total area $V$ minus the total area covered
by the circles $B_{i}$ $,$ i.e., $V$ minus the union $\cup
_{i=1}^{N-1}B_{i}. $ And the formulae (\ref{V}) and (\ref{Vexcl}) do
describe the integral over exactly this area. The formula (\ref{Vexcl}) can
be presented in the form 
\begin{equation}
V_{excl}(x_{1},...,x_{N-1})=\int_{\cup _{i=1}^{N-1}B_{i}}dx_{N}=\mu \left(
\bigcup_{i=1}^{N-1}B_{i}\right) .  \label{V  excl}
\end{equation}%
Indeed, the expression (\ref{Vexcl}) for $V_{excl}$ is exactly the volume of
the union $\cup _{i=1}^{N-1}B_{i}$ known in the set theory, which is
restricted to the intersections of maximum five sets $B_{i}$ and implicitly
excludes core overlapping. Thus, we have expressed the single $N$ th
particle integral in terms of the intersections of the circles $B$ connected
to the other $N-1$ HDs. In the next section we show how this result can be
incorporated in the many-body theory. In order to simplify formulae, in what
follows we omit the explicit presence of the indicator $\theta _{N}$ and
assume that only acceptable configuration are considered whereas all
configurations with any core overlap give zero contribution to $V_{N}.$

\section{Many-body problem}

\subsection{The Speedy-Widom approach: PF is the product of cavities}

To consider implementation of the obtained result in $Z_{N}$ we first follow
Widom's idea \cite{Widom} and transform the PF like that. Divide and
multiply $Z_{N}$ by the PF $Z_{N-1}$ for $N-1$ HDs and introduce the
distribution function (DF) $f_{N-1}$ of $N-1$ disks' coordinates in the
equilibrium system of $N-1$ HDs:%
\begin{equation}
f_{N-1}(x_{1,...,}x_{n-1})=Z_{N-1}^{-1}\exp \left( -\frac{1}{2}%
\sum_{i=1}^{N-1}U_{ij}\right) .  \label{f N-1}
\end{equation}%
Then one has:%
\begin{eqnarray}
Z_{N} &=&Z_{N-1}\int_{\mu
}dx^{N-1}f_{N-1}(x_{1,...,}x_{n-1})V_{N}(x_{1,...,}x_{n-1})  \label{ZN1} \\
&=&Z_{N-1}\left\langle V_{N}\right\rangle _{N-1},  \notag
\end{eqnarray}%
where $\left\langle V_{N}\right\rangle _{N-1}$ is the average volume
accessible for $N$ th disk in the equilibrium system of $N-1$ disks, i.e.,
the cavity $C_{N}$ in a system of $(N-1)$ HDs. Continuing along this line by
introducing the equilibrium distribution functions of a lower and lower
number of HDs, one arrives at the following formula for the PF:%
\begin{equation}
Z_{N}=\left\langle C_{N}\right\rangle _{N-1}\left\langle
C_{N-1}\right\rangle _{N-2}...\left\langle C_{N-N^{\prime }}\right\rangle
_{N-N^{\prime }-1}...V.  \label{ZN2}
\end{equation}%
The result is the product of cavities, the average empty voids in the
equilibrium systems of $N-N^{\prime }-1$ HDs into which the $(N-N^{\prime })$
th HD can be inserted, for all $N^{\prime }$ from $0$ to $N-1.$ This is the
second Speedy's result \cite{Speedy80} which can rightfully be called
Speedy-Widom PF. This result is behind the idea which has been the pivot of
practically all of the search for the equation of states based on the
notions of cavity and spare volume. The problem that has been encountered in
these studies is that, already at liquid densities, cavities become so rare
and finding them in computer simulations so difficult that it was even
dubbed a task futile \cite{1998}. This practically means that in
sufficiently dense HD and hard sphere systems that are still far from their
close packing, $Z_{N}$ is zero, the entropy is minus infinity, and higher
densities are inaccessible because the $N$ th hard core particle cannot be
inserted in such dense systems. This situation is paradoxical as we started
to study a system of $N$ particles but found that one particle has no space
in this system. Below I shall resolve this paradox and derive the consistent
theory in which all HDs have guarantied space in an $N$ HD system.

\subsection{PF is the product of single-disk free volumes}

How could it happen that the rightful guarantied space of $N$ th HD got lost
in a system of $N$ HDs? To answer let us compare the averaging (\ref{ZN1})
of $V_{N}(x_{1},...,x_{N-1})$ with the DF  $f_{N-1}(x_{1},...,x_{N-1})$ of
an equilibrium system of $N-1$ HDs, eq.(\ref{f N-1}), with the integral over 
$V_{N}(x_{1},...,x_{N-1})$ in the second line of eq. (\ref{ZN}). \ The PF  $%
f_{N-1}(x_{1},...,x_{N-1})$ is that in a system of $N-1$ disks and is
established without any effect of an additional disk $N$ of which  $%
f_{N-1}(x_{1},...,x_{N-1})$ never knew. Hence, the factor $V_{N}$ does not
influence $f_{N-1}$, its role is passive and reduces to guiding the external
disk along the maze formed by the $N-1$ \textquotedblright
naitive\textquotedblright\  HDs. In particular, if the maze leaves no place
for an external disk $N$, the integral $\left\langle V_{N}\right\rangle
_{N-1}=0.$  In a dense system, this situation is most probable as the most
probable distribution of $N-1$ disk is uniform. In contrast, the integral (%
\ref{ZN}) over $V_{N}$ is that over $x_{N}$ in a system of $N$ HDs. Now any
collection of $N-1$ disks does know about the presence of another disk which
has the same \textquotedblright naitive\textquotedblright\ status, and is
therefore adjusted as to accomodate it with the probability one. In such
system, the most probable situation is also a uniform distribution, but now
of all $N$ disks (so that $N+1$ st disk could have found no place, but now
this is irrelevant). Thus,  $f_{N-1}$ has \textquotedblright no
idea\textquotedblright\ of the $N$ th disk whereas all the $N-1$
coordinates, the arguments of $V_{N}$ in (\ref{ZN}), do\ keep knowledge of
the $N$ th disk to which they cannot approach to a distance below $\sigma $.
To summarize, in the PF (\ref{ZN2}), the average of $V_{N}$ in the original
statistical integral was replaced by a different average (\ref{ZN1}) with
the equilibrium DF for the different system. Hence we have to base our
theory on the integral $V_{N}.$

The integral $V_{N}$ defined in (\ref{V}) depends on the coordinates of all $%
N$ disks, explicitly on $x_{1},x_{2},...,x_{N-1}$ and implicitly, via these $%
N-1$ coordinates, on $x_{N}$. However, it is not difficult to see that, in
the thermodynamic limit$,$ this integral tends to a constant value which
does not depend on all the $N$ coordinates. To see this let us divide the
infinite volume $V$ into, e.g., $\sqrt{N}$ equal subvolumes $\Delta V_{i}$
of size $V/\sqrt{N}$ with $N/\sqrt{N}=\sqrt{N}$ disks in each and the
density $\rho =\sqrt{N}/(V/\sqrt{N})=N/V$. The system of HDs (and hard
spheres) is not only ergodic, but possesses a mixing property \cite%
{Sinai1,2,KSF} which implies in particular that, although distributions of
disks in different $\Delta V_{i}$ are same, the actual arrangements of disks
in different $\Delta V_{i}$ are different. In other words, the disks
arrangements in different infinite $\Delta V_{i}$ represent an infinite
number of different realizations of distributions of same number of disks
and density $N/V$ in the similar infinite size systems (both $N/\sqrt{N}$
and $V/\sqrt{N}$ are infinite). Then it follows that the integral $V_{N}$
over the volume $V$ is the thermodynamic average over infinite ensemble of
realizations, i.e., is a constant $\left\langle V_{N}\right\rangle _{N}$
which depends only on the density $N/V.$ By definition, $\left\langle
V_{N}\right\rangle _{N}$ is the thermodynamic limit of the free volume of a
single disk in the equilibrium system of density $N/V.$ Similarly, defining
the one-particle integral in the system of $N-N^{\prime }$ HDs, $0\leq
N^{\prime }\leq N-1,$ we obtain the thermodynamic limit $\left\langle
V_{N-N^{\prime }}\right\rangle _{N-N^{\prime }}$ of a single-particle free
volume in the system of $N-N^{\prime }$ particles. Presenting the PF in the
"factorized" form and continuing this process, in the thermodynamic limit we
obtain the PF in the following form:%
\begin{eqnarray}
Z_{N} &=&\prod\limits_{k=2}^{N}\int_{V}dx_{k}\exp \left( -\frac{1}{2}%
\sum_{i=1}^{k-1}U_{ki}\right)   \notag \\
&\rightarrow &\left\langle V_{N}\right\rangle _{N}\left\langle
V_{N-1}\right\rangle _{N-1}...\left\langle V_{2}\right\rangle _{2}V.
\label{ZZ}
\end{eqnarray}%
This PF is the product of the average free volumes of a single particle in
the equilibrium systems of $N$, $N-1,...$ HDs and is essentially different
from the Speedy-Widom PF (\ref{ZN1}). The average free volume $\left\langle
V_{N-N^{\prime }}\right\rangle _{N-N^{\prime }}$ comprises both average
private cell $\left\langle c_{N-N^{\prime }}\right\rangle _{N-N^{\prime }}$
and the average cavity $\left\langle C_{N-N^{\prime }}\right\rangle
_{N-N^{\prime }}$ and both are obtained in the equilibrium system of $%
N-N^{\prime }$ HDs, i.e., not reduced by one. The pressure can be computed
as $P_{N}\propto -\partial \ln Z_{N}/\partial V=-N\partial \ln
Z_{N}/\partial \rho $. To complete our task, in the next section we connect
the free volume $\left\langle V_{N-N^{\prime }}\right\rangle _{N-N^{\prime }}
$ with the multiple intersections of $N-N^{\prime }$ disks of radius $\sigma 
$ in the equilibrium system of the same number $N-N^{\prime }$ of disks.

\section{Relation between single-disk free volume and multiple disks'
intersections}

\subsection{The total excluded and free volume for a single HD.}

The free volume of a disk in an $N$ disk system is the cavity left by the
rest $N-1$ disks. We see that the practical definition of the free volume is
related to removing $N$ th disk from the $N$ HD system to which it belongs.
However, as we showed above, dealing with such objects one should be
careful.  Therefore, it is both convenient and essential to relate the
expression for the free volume in a system of $N$ disks in terms of the
equilibrium system of this very number of disks $N$. Moreover, the PF (\ref%
{ZZ}) makes this task crucial.

The formula (\ref{V}) for the free volume of $N$ th disk in a system of $N$
disks is correct but inconvenient because it is related to the reduced
distribution function of $N-1$ disks which is highly nontrivial. Let us
consider instead directly the system of $N$ disks. Assume that center of
disk $N$ is at $x^{\prime }$ and let us find the free volume for this disk.
Above we noted that the division on cavity and private cell depends on $%
x^{\prime }$ but their sum, which is what we actually need, does not. Our
task is thus to find this sum in terms of all $N$ disks' intersections. This
sum is represented by the total white area in Fig.1, i.e., the cavity left
by the rest $N-1$ disks, but in the $N$ disk system! The excluded area of
points $x_{N}\in V$ created by all $N$ disks, $\bigcup_{i=1}^{N}B_{i},$
exceeds the excluded area due to the rest $N-1$ disks by the area of the $%
\sigma $ circle $B_{N}$, but without all its areas $B_{N}\cap B_{i}$ already
covered by other $N-1$ disks (because these areas should not be counted
twice), Fig.1. Thus, the single disk excluded volume $V_{N,excl}$, where $%
x^{\prime }$ cannot enter, is the integral (\ref{V excl}) over this
inaccessible area: 
\begin{equation}
V_{N,excl}=\mu \left( \bigcup_{i=1}^{N}B_{i}\right) -\pi \sigma ^{2}+\mu
\left( \bigcup_{i=1}^{N-1}(B_{N}\cap B_{i})\right) .  \label{VVexcl}
\end{equation}%
Making use of eq.(\ref{V}) \ and (\ref{Vexcl}), one obtains the formula
which expresses the free volume of a single disk $i_{0}$ in a system of $N$
HDs via multiple intersections of $N$ $\sigma $ circles:%
\begin{eqnarray}
V_{N} &=&V-\left[ \sum\limits_{i=1}^{N}\left( \mu _{i}+\sum_{i>j}^{N}\mu
_{ij}-\sum_{i>j>k}^{N}\mu _{ijk}+\sum_{i>j>k>l}^{N}\mu
_{ijkl}-\sum_{i>j>k>l>m}^{N}\mu _{ijklm}\right) \right.   \notag \\
&&\left. -\left( \pi \sigma ^{2}-\sum_{i}\mu _{i_{0}i}+\sum_{j>k}\mu
_{i_{0}jk}-\sum_{j>k>l}\mu _{i_{0}jkl}+\sum_{j>k>l>m}\mu _{i_{0}jklm}\right)
_{i,j,k,l,m\neq i_{0}}\right] .  \label{VVV}
\end{eqnarray}%
This formula shows that the total free volume for the disk $i_{0}$ is equal
to the cavity in the $N$ disk system plus $\pi \sigma ^{2}$ minus the area
of intersection of disk $i_{0}$\ with the rest $N-1$ disks. Both formulae (%
\ref{VVexcl}) and (\ref{VVV}) do not refer to any system of $N-1$ HDs: the
upper summation limit $N-1$ in the excluded volume in the form (\ref{V excl}%
) is replaced by $N$, the second term in (\ref{VVexcl}) also determines the
intersection areas in the $N$ HD system as indicated by the presence of $%
B_{N}$. In the thermodynamic limit\ the above $V_{N}$ tends to the constant $%
\left\langle V_{N}\right\rangle =\lim_{N,V\rightarrow \infty }V_{N}$ which
is the function of $N/V.$ It can be expressed in terms of the following
average values $v_{N,n}$ which are different intersections of an individal $%
\sigma $ circle averaged over all $\sigma $ circles: 
\begin{eqnarray}
\mu _{i} &=&\pi \sigma ^{2},  \notag \\
\frac{1}{N}\sum\limits_{i=1}^{N}\left( \sum_{j\neq i}^{N}\mu _{ij}\right) 
&=&v_{N2},  \notag \\
\frac{1}{N}\sum\limits_{i=1}^{N}\left( \sum_{j>k}^{N}\mu _{ijkN}\right)
_{i\neq j,k} &=&v_{N3,}  \label{v} \\
\frac{1}{N}\sum\limits_{i=1}^{N}\left( \sum_{j>k>l}^{N}\mu _{ijklN}\right)
_{i\neq j,k,l} &=&v_{N4},  \notag \\
\frac{1}{N}\sum\limits_{i=1}^{N}\left( \sum_{j>k>l>m}^{N}\mu _{ijklm}\right)
_{i\neq j,k,l,m} &=&v_{N5},  \notag
\end{eqnarray}%
where index $N$ indicates that the average is computed in an $N$ HD system
and another index indicates the number of intersecting $\sigma $ circles. We
remember that for a fixed $i$, the maximum number of terms in the above sums
is five so that the summations are actually not extensive. In terms of
simulation results, the procedure of finding $v_{N,n}$ consists of computing
all intersection of each disk with other $n-1$ $\sigma $ circles and then
averaging over the results. It is essential that the areas of all
intersections of our interest are uniquely determined by the coordinates of
the participating disks and can be computed analytically making use of the
formulas derived in \cite{Kratky1,Kratky2,Sheraga}. The detailed computation
of the intersection volumes $v_{Nn}$ in the densely packed triangular HD
lattice is presented in section 5.

\subsection{Extensive and intensive free volume terms: the cavity and the
private cage cell}

Separating the extensive and intensive terms in (\ref{VVV}) in the context
of (\ref{v}), one finally obtains: 
\begin{equation}
\left\langle V_{N}\right\rangle =V_{N,exten}+V_{N,inten},  \label{VNbar}
\end{equation}%
where 
\begin{equation}
V_{N,exten}=\left\langle C_{N}\right\rangle _{N}=V-N\left( \pi \sigma
^{2}-v_{N2}/2+v_{N3}/3-v_{N4}/4+v_{N5}/5\right) ,  \label{Vext}
\end{equation}%
\begin{equation}
V_{N,inten}=\left\langle c_{N}\right\rangle _{N}=\pi \sigma
^{2}-v_{N2}+v_{N3}-v_{N4}+v_{N5}.  \label{Vint}
\end{equation}%
The denominators in $V_{N,exten}$ reflect the fact that in the sum over all
disks, any intersection of $n$ circles is counted $n$ times; at the same
time, the intersections of a single circle in $V_{N,inten}$ are counted only
once. The free volume contains two contributions, the extensive $%
V_{N,exten}=\left\langle C_{N}\right\rangle _{N},$ which is the average
cavity volume in the $N$ HD system, and the intensive $V_{N,inten}=\left%
\langle c_{N}\right\rangle _{N},$ which is the guarantied volume of a cell
connected to or, better to say, containing a single HD. This cell of size $%
\left\langle c_{N}\right\rangle _{N}$ is available for any single disk even
if the cavity practically vanishes, which is the case of high densities. We
say practically because in an infinite system a fluctuation in the form of
cavity, however small its probability be, still must exist. This situation
is expected to be similar to that with the so-called windowlike defects in a
quasi-one-dimensional HD system \cite{WE}: in the thermodynamic limit, the
probability to have such a "window" in the crystalline zigzag vanishes only
at the close packing \cite{JCP2020,WE,Saika}. Coming back to our
two-dimensional HD system, we see that at densities close to crystallization
densities, the intensive term, which is fully negligible at lower densities
where cavity dominates, becomes the only source of entropy. In that case $%
\left\langle V_{N}\right\rangle $ is the volume of a cell, $\left\langle
c_{N}\right\rangle _{N},$ in which a single HD is caged by its neighbors.
Note that the size of this cell, eq.(\ref{Vint}), is determined not only by
the central disk's next neighbors but also by next next neighbors that can
contribute to the intersection if their centers are within the $\sigma $
circle of the central disk. Such intersections with the next and next next
neighbors has been taken into account, at a phenomenological level, in the
cell models of the equation of state for HD systems \cite%
{1968,1972,1979,PRE2006}. It is this cell's size that was the main task of
the cell models: the counterpart of $V_{N,inten}$ was computed in
configurations that were assumed to contribute the most (usually these were
the symmetric configurations related to the triangular lattice). Our
formulas show the way to find the cell size as thermodynamic average.
Moreover, formulas (\ref{Vext}) and (\ref{Vint}) give the total free volume
in an $N$ HD system so that a) the cavities and cells are not considered
separately, b) they can be computed directly from disks' coordinates even
analytically, and c) only the original, the very same $N$ HD system needs to
be considered. Thus, the idea of both extensive and intensive free volume
contributions put forward by Hoover and Ree \cite{1968} and Hoover, Ashurst,
and Grover \cite{1972} has been embodied in our theory unifying both terms
in the frame work of the disks' intersection device first pointed to by
Speedy \cite{Speedy80}. In the next section I present an example of
analytical calculation of both $\left\langle C_{N}\right\rangle $ and $%
\left\langle c_{N}\right\rangle _{N}$ at densely packed triangular lattice
which shows that the two terms in this state vanish as expected. This
demonstrates their independence and different status.

\section{Vanishing of the cavity and intensive cell volume in a close packed
triangular lattice}

Here the procedure of counting and computing areas of all possible multiple
intersections of the $\sigma $ circles for a single disk and computing the
cavity and cell volume is demonstrated for a close packed triangular lattice
which has the single configuration and does not need averaging. A fragment
of this lattice is shown in Fig.2. Disks' cores, which are in contact with
each other, are filled and have radius $\sigma /2;$ the attached concentric $%
\sigma $ circles of radius $\sigma $ are shown by dashes (shown only for
five disks). The central disk $0$ is shown along with its six next
neighbors, $1,2,3,4,5,6,$ and with its six next next neighbors $%
7,8,9,10,11,12$; the distance of these next next neighbors to $0$ is less
than $\sigma $ so that their $\sigma $ circles can overlap with the central $%
\sigma $ circle. The upper fragment with the five shown $\sigma $ circles is
sufficient for the establishing of all neighbor $\sigma $ circles
overlapping with the central circle because it is one of the three similar
fragments. No other disks in the lattice have their $\sigma $ circles
overlapping with $0$ circle. First we notice that no five circles intersect
in this lattice. Next, for the circles indicated by dashes, we find and list
different $\sigma $ pairs, $\sigma $ triples, $\sigma $ quadruples, which
include $0$ circle; then we compute their surface areas, and count the total
numbers of such different terms, and finally use the formulae (\ref{Vint})
and (\ref{Vext}). 
\begin{figure}[tbp]
\includegraphics[height=12cm]{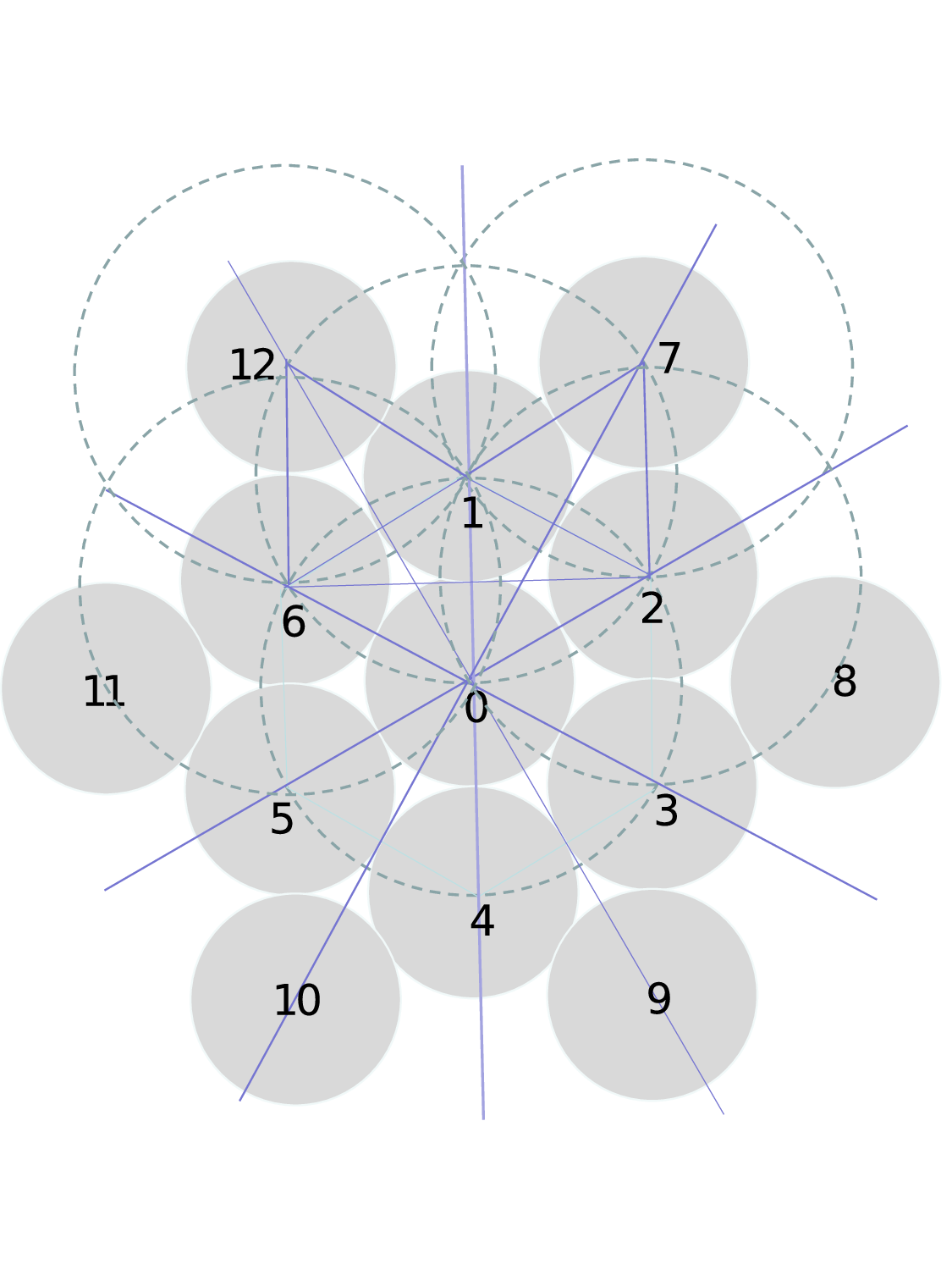}
\caption{{}A fragment of triangular densely packed HD lattice. The disks'
cores are filled circles of the radius $\protect\sigma /2.$ Five $\protect%
\sigma $ circles centered at $0,$ 2, 7$,$ 12, and 6 are indicated by dashes.
The central disk $0$ has next neighbors centered at 1,2,3,4,5, and 6, and
next next neighbors centered at 7,8,9,10,11, and 12. \ $\protect\sigma $
circles of all disks in the lattice, which are not shown, do not overlap
with the $\protect\sigma $ circle centered at $0.$}
\label{Fig2}
\end{figure}

\textit{Surface area of two disk intersection} $S_{2}.$ There are six pairs
of circles similar to $01$ and six pairs similar to $07,$ which gives for
the total contribution of pairs $S_{2}=6S_{01}+6S_{07}.$ The $S_{01}$ is the
area $0216$ bounded by circle 0 from above and circle 1 from below, its area
is $S_{01}=2(\pi /3-\sqrt{3}/4)\sigma ^{2};$ the area $S_{07}$ is the lobe
with the vertices $1$ and 2, $S_{07}=2(\pi /6-\sqrt{3}/4)\sigma
^{2}:S_{2}/\sigma ^{2}=6(\pi -\sqrt{3}).$

\textit{Surface area of three disk intersection} $S_{3}.$ Similarly, $%
S_{3}=6S_{102}+12S_{017}+6S_{602}=6S_{102}+18S_{017}$ as $S_{602}=S_{017}.$
The area $S_{102}$ is that of the curvilinear triangle $102$, $S_{102}=[%
\sqrt{3}/4+3(\pi /6-\sqrt{3}/4)]\sigma ^{2};$ another triple area $S_{017}$
is that of the lobe with the vertices $1$ and 2, so that $%
S_{017}=S_{07}=2(\pi /6-\sqrt{3}/4)\sigma ^{2}:S_{3}/\sigma ^{2}=-48\sqrt{3}%
/4+9\pi .$

\textit{Surface area of four disk intersection} $S_{4}.$ $%
S_{4}=6S_{0216}+6S_{0172}=12S_{0216}.$ Finally, the area $S_{0216}$ is the
lobe with vertices 0 and 1 which is equal to $S_{07},$ $S_{0216}=2(\pi /6-%
\sqrt{3}/4)\sigma ^{2}:S_{4}=4\pi -6\sqrt{3}.$

Now we are ready to compute the close packing values of the intensive cell
volume $c_{cp}=V_{cp,int}$ and the cavity $C_{cp}=V-V_{cp,ext}$ using the
formulae (\ref{Vint}) and (\ref{Vext}). Substituting the above values of $%
S_{n},$ one finds: 
\begin{equation}
c_{cp}/\sigma ^{2}=\pi -\left[ 6(\pi -\sqrt{3})-(-48\sqrt{3}/4+9\pi )+4\pi -6%
\sqrt{3}\right] =0.  \label{c}
\end{equation}%
The cavity is extensive and to deal with the size independent quantities, we
divide $V-V_{N,exten}$ by $N\sigma ^{2}.$ The expression $V/N\sigma ^{2}=\pi
/(4\eta _{cp})$ where $\eta _{cp}=N\pi \sigma ^{2}/4V$ is the packing
fraction at close packing, $\eta _{cp}=\pi /2\sqrt{3}$. Substituting the
above values of $S_{n}$ in $V_{pc,exten}$ one gets:%
\begin{equation}
C_{cp}/N\sigma ^{2}=\pi /4\eta _{cp}-\pi +\pi -\sqrt{3}/2=0.  \label{C}
\end{equation}%
It is essential that not only the total free volume, but both intensive and
extensive free volumes vanish separately which shows their functional
independence. It is also important to realize that, as evident from Fig.2, a
small increase in disks' separation will results in a nonzero intensive $c$
whereas the extensive $C$ will remain zero until at some density, which
might be close to that of crystallization, it will start to grow. As each
factor in the PF $Z_{N}$ (\ref{ZZ}) corresponds to different density $%
(N-N^{\prime })/V$, this point will be appearing consequentially in the
factors $\left\langle V_{N-N^{\prime }}\right\rangle $ with progressively
lower $N-N^{\prime }.$ Can this process result in a discontinuity of $Z_{N}$%
? This is one of the questions the method presented in this paper is
expected to answer.

\section{Concluding remarks}

Preliminary results for the free volumes and entropy of a two-dimensional HD
system, calculated by the formulae of this paper, have been recently
obtained from a molecular dynamics simulation \ \cite{MD}. The results show
robustness of this method and its ability to pick main peculiarities of the
phase behavior of a two-dimensional HD system. The work is in progress.

In conclusion I would like to speculate about possible implication of the
results obtained in this paper for an analytical approach. Our results show
that the entropy can be computed provided the four functions of the system
density $\rho $ are known, i.e., $v_{N2}(\rho ),v_{N3}(\rho ),v_{N4}(\rho ),$
and $v_{N5}(\rho ).$ The simulations can give one an idea about these $\rho $
dependences which can advance our "analytical" understanding of the
two-dimensional HD system. The formulae obtained in this paper are equally
applicable for a three-dimensional system of hard spheres. The main
difference is that, in the last case, the computation are expected to be
much more extensive\ as one will need to deal with the intersection of up to
eleven $\sigma$ spheres allowed without their hard core overlap.

\section*{Acknowledgements}

Author is greatful to T. Bryk and A. Trokhymchuk for showing their preliminary results of molecular dynamic simulation
 and  to Center for Theoretical Physics PAS for hospitality. This research is part of the project No.
2022/45/P/ST3/04237 co-funded by the National Science Centre and the
European Union Framework Programme for Research and Innovation Horizon 2020
under the Marie Sk\l odowska-Curie grant agreement No. 945339.

\end{document}